\documentstyle{article}
\begin{document}
\begin{center}
{\bf Radiative capture on neutron rich nuclei}

Karl Gr\"un, Rudolf Pichler, Heinz Oberhummer\\
\vspace{10pt}
Institut f\"ur Kernphysik, TU Wien,
A--1040 Vienna, Austria
\end{center}
\begin{center}
{\bf ABSTRACT}
\end{center}
The $({\rm n},\gamma)$--cross sections for neutron--rich oxygen isotopes have
been calculated in the direct capture model. The experimental data for
${^{18}{\rm O}}({\rm n},\gamma){^{19}{\rm O}}$ can be reproduced using this
model.
 Compared to
previous work the cross section for
${^{19-21}{\rm O}}({\rm n},\gamma){^{20-22}{\rm O}}$ are enhanced considerably
by factors between four and 300.
\begin{center}
{\bf 1.~INTRODUCTION}
\end{center}

In the last years the importance of the direct--reaction
(DI) mechanism in nucleosynthesis has been realized.
The DI may even dominate over the
compound--nucleus (CN) reaction mechanism in nuclear processes relevant
for nucleosynthesis.
This can be the case for primordial and stellar nucleosynthesis, f.~i.~in
neutron--induced
reactions in the inhomogenious big--bang scenario$^{1-3}$, in many solar
nuclear reactions$^4$, in the famous triple--alpha process leading to the
creation of carbon$^{5}$, or in nuclear reactions of the r--process
involving magic--shell nuclei and nuclei far from stability$^6$.

In this work we calculate cross sections of neutron--capture on
oxygen isotopes. The reaction rates for these reactions are of
importance in the inhomogeneous big--bang scenario$^{7,8}$
as well as in the $\alpha$--rich freeze out of the
neutron--rich hot neutrino bubble in supernovae
type II$^{9,10}$.
We investigate the capture reaction $^{18}$O(n,$\gamma$)$^{19}$O
at thermal (0.025\,eV) and thermonuclear (30\,keV) projectile energies
and compare the calculated cross sections with the experimental data.
Using the same methods   we also calculate the cross sections for
the neutron--capture reactions $^{19-21}$O(n,$\gamma$)$^{20-22}$O
in the thermal and thermonuclear
energy range.

In  section 2 the folding procedure for the determination of the
optical and bound--state potentials is introduced. In section 3 we discuss
the model necessary for the calculation of direct capture (DC) cross sections.
The method of bound states embedded into continuum (BSEC) allows including
isolated
CN--resonances in the DC--cross section. In section 4 the calculations
and results for the neutron capture cross sections for some neutron--rich
oxygen isotopes are presented and discussed. Finally, in the last section a
short summary is given.
\begin{center}
{\bf 2.~FOLDING PROCEDURE}
\end{center}

The folding procedure is used for calculating the
nucleon--nucleus potentials in order to describe the
elastic scattering data and the bound states. This method was already applied
successfully in describing many nucleon--nucleus systems.
In the folding approach the nuclear density
$\rho_{A}$ is folded with an energy and density dependent NN interaction
$v_{\rm eff}$$^{11,12}$
\begin{equation} \label{1}
V(R) = \lambda V_{\rm F}(R) = \lambda \int
\rho_{A} (\vec{r}) v_{\rm eff}(E,\rho_A,|\vec{R}  - \vec{r}|)
d\vec{r}
\end{equation}
with $\vec{R}$ being the separation of the centers of mass of the two colliding
nuclei. The normalization factor $\lambda$ is adjusted to elastic
scattering data
and to bound-- and
resonant--state energies. The potential obtained in
this way ensures the correct behavior of the wave functions in the nuclear
exterior.
At the low energies considered in nucleosynthesis the
imaginary parts of the optical potentials are small and can often be neglected.

In the folding approach the nuclear densities $\rho_{A}$ for stable nuclei
are derived from experimental charge distribu\-tions$^{13}$. For unstable
nuclei
these densities are determined with the help of the relativistic mean
field theory (RMFT).
The RMFT describes the nucleus as a system of Dirac nucleons interacting via
various meson fields. In the last few years this theory has turned out to be a
very successful tool for the description of many nuclear properties (for
example binding energies and charge radii for stable isotopes)$^{6,14}$.

\begin{center}
{\bf 3.~REACTION MECHANISMS AND MODELS}
\end{center}

In nuclear reactions two extreme types of reaction mechanisms can exist:
the compound--nucleus (CN) and the direct (DI) process. In the CN mechanism
the projectile merges in the target
nucleus and excites many degrees of freedom of the CN. The excitation proceeds
via a multistep process and therefore has a reaction time typically of the
order $10^{-16}\,$s to $10^{-20}\,$s. After this time the CN decays into
various exit channels. The relative importance of the decay channels is
determined by the branching ratios to the final states.
In the DI process the projectile excites only a
few degrees of
freedom (e.g.~single--particle or collective). The excitation proceeds in one
single step and has a characteristic time scale of $10^{-21}\,$s to
$10^{-22}\,$s. This corresponds to the time the projectile needs
to pass through the target nucleus; this time is much shorter than the
reaction time of CN processes.

In thermonuclear scenarios the projectile energy is well below the Coulomb
and/or centrifugal barrier. At these energies the competition between
different reaction
mechanisms is quite complicated. At these
energies the CN formation may be suppressed, because there may exist no
CN levels that can be populated, especially in light, magic and
far--off--stability nuclei.

In this work the theoretical cross sections are given through the
 DC--contribution.
The theoretical cross section $\sigma^{\rm th}$
is obtained from the
DC cross section $\sigma^{\rm DC}$ by$^{4,15}$
\begin{equation} \label{2}
\sigma^{\rm th} = \sum_{i} \: C_{i}^{2} S_{i}\sigma^{\rm DC}_{i} \quad .
\end{equation}
The sum extends over the ground state and
excited states in the final nuclei, where the spectroscopic
factors $S_{i}$ are known. The isospin Clebsch--Gordan coefficients
are given by $C_{i}$. The DC cross sections $\sigma^{\rm DC}_{i}$ are
essentially
determined by the overlap of the scattering wave function
in the entrance channel, the bound--state wave function
in the exit channel and the multipole transition--operator. The radial
dependence
of the DC--integral is in our case determined uniquely by the folding
potentials.

Isolated CN--resonances with known resonance energy and width can be
incorporated
into the nonresonant DI cross section by BSEC$^{16-18}$
which allows a simultaneous calculation of the resonant and
non--resonant contributions of the cross sections. In this
approach the scattering wave function is calculated using
an adequate energy--dependent potential generating
single--particle resonances and reproducing effects caused
by their coupling with complicated bound states
into the continuum.
\begin{center}
{\bf 4.~CALCULATIONS AND RESULTS}
\end{center}

The density distributions of $^{18}$O to $^{22}$O
were derived with the help of RMFT.$^{14}$ In the entrance
channel the normalisation
factor $\lambda$ given in  Eq.~\ref{1} of $^{18}$O+n
was adjusted to reproduce the total cross section.$^{19}$
For the other reactions the $\lambda$ were adjusted
to reproduce the same volume integral
$J = 578.33 \thinspace {\rm MeV}\thinspace {\rm fm}^{3}$ as for the folding
potential for $^{18}$O+n elastic scattering.
In the exit channels the $\lambda$ were fixed by
the neutron separation energy from the final nuclei.
The folding potentials of Eq.~\ref{1} were determined with the help of the
computer code DFOLD.$^{20}$

The spectroscopic factors for one--nucleon transfer
of the oxygen isotopes were
determined from shell--model calculations in the
sd--shell with an effective
nucleon--nucleon interaction derived by Wildenthal$^{21}$. For these
calculations the programs GENESIS, RITSSCHIL, DIA$^{22}$
and SPECTROS$^{23}$ were used.
The resonance energies and widths of $^{19}$O used for the BSEC--calculations
of $^{18}$O(n,$\gamma$)$^{19}$O are listed in Ref.~24.
For the direct--capture calculations Eq.~\ref{2} the code TEDCA$^{25}$ was
used.
For the BSEC--calculations the code TEDCA was generalized
to include this approach.

Let us first discuss the reaction $^{18}$O(n,$\gamma$)$^{19}$O.
For this reaction experimental reaction data at the
thermal energy (0.025\,meV) and around the ${3/2}^-$-resonance
at 625\,keV are available.
The value of the cross section at 23.3\,keV is given
by $\approx 8 \pm 1$\,$\mu$barn$^{26,27}$.
We consider this reaction as a test for our DC--model.
\begin{figure}
\vspace{8cm}
\caption{ Level scheme for the reaction $^{18}$O(n,$\gamma$)$^{19}$O and
comparison of the DC cross section for $^{18}$O(n,$\gamma$)$^{19}$O
with the experimental data$^{24,26,27}$.}
\end{figure}

The level scheme for this reaction is shown on the left side in Fig.~1.
There are two types of E1--transitions relevant for
the energy range between thermal and thermonuclear
energies. The first one is from an s--wave in the
entrance channel exciting the negative--parity state
3/2$^{-}$ in the final nucleus which
is just bound by about 12.1\,keV. Even so this transition
has a very low Q--value it still reproduces the experimental
cross section $\sigma_{\rm therm} = 0.16 \thinspace {\rm mb}$ at
0.025\,eV with a reasonable spectroscopic factor of $0.331$.
In this case the transition matrix  has
contributions  up to 500\,fm and has a maximum
at about 100\,fm. This transition gives the well--known
1/v--behavior (see Fig.~1). The second type of transition comes from
an initial p-wave and excites the positive--parity states 5/2$^+$, 3/2$^{+}$,
1/2$^{+}$ in the final nucleus. These transitions
have a v--behavior (see Fig.~1). Furthermore as can  be seen from Fig.~1,
the resonance 3/2$^{-}$ at 625\,keV can be described by the BSEC which
was discussed before.
\begin{table}
\caption{Comparison of the contributions of different
transitions to bound--states and their spectroscopic factors
between previous work$^{8}$ and this work for the
reactions $^{18-21}$O(n,$\gamma$)$^{19-22}$O}
\begin{center}
\begin{tabular}{||ccccrr||}
\hline
reaction & final state & spectroscopic factor & transition  & previous
work$^{8}$ &
          this work \\
         & & &  &
	  \multicolumn{2}{c||}{$S_{\rm n} \thinspace [\mu{\rm b}/\sqrt{\rm MeV}]
	     \thinspace (E = 30 \thinspace {\rm keV})$ } \\
  \hline
${^{18}{\rm O}}({\rm n},\gamma){^{19}{\rm O}}$
         & ${5/2}^+$ & 0.687  &  $p \rightarrow d$  & &   $8.85$
\hspace{0.5 cm}    \\
	 & ${3/2}^+$ & 0.013  &  $p \rightarrow d$  & &   $0.13$  \hspace{0.5 cm}
  \\
	 & ${1/2}^+$ & 0.830  & $p \rightarrow s$   & &   $44.43$ \hspace{0.5 cm}
\\
	 \cline{4-6}
	 &           &
	    & total          &    $52.40^{1}$ \hspace{10 pt} & $53.33$ \hspace{0.5
cm} \\
	 \hline
${^{19}{\rm O}}({\rm n},\gamma){^{20}{\rm O}}$
         & $0^+$     & 3.427  &  $p \rightarrow d$ & $0.54$ \hspace{0.5 cm}
&   $2.72$ \hspace{0.5 cm}   \\
         & $2^+$     & 0.731  &  $p \rightarrow d$ & $1.94$ \hspace{0.5 cm}
&   $2.63$ \hspace{0.5 cm}   \\
         &           & 0.142  &  $p \rightarrow s$ & $4.15$ \hspace{0.5 cm}
&   $5.16$ \hspace{0.5 cm}   \\
         & $4^+$     & 1.021  &  $p \rightarrow d$ & &   $4.63$ \hspace{0.5
cm}    \\
         & $2^+$     & 0.574  &  $p \rightarrow s$ & &   $15.58$
\hspace{0.5 cm}   \\
	 \cline{4-6}
	 &           &       & total
	    & $6.63$ \hspace{0.5 cm} & $30.72$ \hspace{0.5 cm} \\
	 \hline
${^{20}{\rm O}}({\rm n},\gamma){^{21}{\rm O}}$
         & ${5/2}^+$ & 0.345  &  $p \rightarrow d$ & $11.02$ \hspace{0.5
cm} &  $5.40$ \hspace{0.5 cm} \\
	 & ${1/2}^+$ & 0.811  &  $p \rightarrow s$ & &   $40.23$ \hspace{0.5 cm}
   \\
	 \cline{4-6}
	 &           &       & total
	    & $11.02$ \hspace{0.5 cm} & $45.63$ \hspace{0.5 cm}  \\
	 \hline
${^{21}{\rm O}}({\rm n},\gamma){^{22}{\rm O}}$
         & $0^+$ & 5.222  &   $p \rightarrow d$ & $0.16$ \hspace{0.5 cm} &
 $4.81$ \hspace{0.5 cm} \\
         & $2^+$ & 0.821  &   $p \rightarrow s$ &  &   $22.42$ \hspace{0.5
cm} \\
         & $3^+$ & 0.771  &   $p \rightarrow s$ &  &   $21.44$ \hspace{0.5
cm} \\
	 \cline{4-6}
	 &           &       & total
	     & $0.16$ \hspace{0.5 cm} & $48.67$ \hspace{0.5 cm} \\
	 \hline
\end{tabular}
\end{center}
{\footnotesize $^1$ Determined with the help of the experimental value of
the cross
section at 23.3 keV.}
\end{table}

The contributions of the different transitions to the cross section
at 30\,keV and their sum for the $^{18-21}$O(n,$\gamma$)$^{19-22}$O reactions
are shown in table 1. Also the spectroscopic factors obtained from
the shell--model calculations are listed. One can see, that
the cross sections  calculated in this work
are enhancend by factors between four and 300.
The reasons are that in previous work$^{8}$
spectroscopic factors have been assumed
too low and important transitions to some excited
final states have been neglected.

\begin{center}
{\bf 5.~SUMMARY}
\end{center}

Direct--capture calculations using the folding procedure
can reproduce excellently the experimental data
for neutron--rich nuclei in the thermal as well as in thermonuclear
energy region. Isolated resonances can also be incorporated sucessfully
in the
direct capture model. The cross sections calculated in our model
for neutron--rich oxygen isotopes are
for some reactions considerably
higher than the ones given previously. Such changes should
lead to  modifications of existing reaction network
calculations.
\begin{center}
{\bf 6.~ACKNOWLEDGMENTS}
\end{center}

We want to thank the  Fonds zur F\"orderung der
wissenschaftlichen Forschung in \"Osterreich (project P8806--PHY)
for their support. We are also oblidged to H.~Beer and J.~Meissner
for valuable and helpful discussions concerning the experimental data and
to H.~Herndl concerning the spectroscopic factors.
\begin{center}
{\bf 7.~REFERENCES}
\end{center}

1. T.~Rauscher, H.~Krauss, K.~Gr\"un, H.~Oberhummer and E.~K\'wasniewicz,
"Analysis of  $^{8}$Li($\alpha$,n)$^{11}$B below the Coulomb barrier in
the Potential Model", {\it Phys.~Rev.\/}, Vol.~C45, pp.1996--2000, 1992.\\
\indent
2. C.E.~Rolfs, W.S.~Rodney, "Cauldrons in the Cosmos", University of
Chicago Press, Chigaco, 1988.\\
\indent
3.~H.~Krauss, K.~Gr\"un, T.~Rauscher and H.~Oberhummer, "The Astrophysical
S--factor of the reaction $^{7}$Be(p,$\gamma$)$^{8}$B in the Direct
Capture Model", {\it Ann.~Physik\/}, Vol.~2, pp.~256--264, 1993.\\
\indent
4.~H.~Oberhummer and G.~Staudt: "Direct mechanism in solar nuclear
reactions", Editor: H.~Haubold,{\it Proceedings of the
Third United Nations/European Space Agency Workshop on Basic Space
Science for Developing Countries},  Lagos, Nigeria,
18--22 October 1993; {\it AIP Conference Proceedings\/}, in press.\\
\indent
5.~P.~Mohr, H.~Abele, V.~K\"olle, G.~Staudt and H.~Oberhummer:
"Properties of $^{8}$Be and $^{12}$C deduced from the folding--potential
model",
{\it Z.~Physik A\/}, in press.\\
\indent
6.~W.~Balogh, R.~Bieber, H.~Oberhummer, T.~Rauscher, K.-L.~Kratz,
P.~Mohr, G.~Staudt and M.M.~Sharma, "Direct capture at low energies",
{\it Proceedings of the European Workshop on Heavy Element Nucleosynthesis,
Budapest, March 9--11 1993\/}, Institute of Nuclear Research of
the Hung.~Acad.~of Sci., S.67--73.\\
\indent
7.~R.~Malaney and G.J.~Mathews, "Probing the early universe:
  A review of primordial nucleosynthesis beyond the standard big bang",
  {\it Phys.~Rep.\/}, Vol.~ 229, pp.~145--229, 1993.\\
\indent
8.~T.~Rauscher, J.H.~Applegate, J.J.~Cowan, F.-K.~Thielemann and
M.~Wiescher, "Production of Heavy Elements in Inhomogenious Cosmologies"
{\it ApJ\/}, in press.\\
\indent
9.~W.M.~Howard, S.~Gloriely, M.~Rayet and M.~Arnould,
"Neutron--rich $\alpha$--rich freeze--out and the r--process
in the high entropy neutrino--energized supernova bubble",
{\it ApJ\/}, Vol.~417, pp.~713--724, 1993.\\
\indent
10.~J.~Witti, H.-Th. Janka and K.~Takahashi, "Nucleosynthesis
in neutrino--driven winds from protoneutron stars --- I\@.
the $\alpha$--process", {\it Astronomy and Astrophysics}, submitted.\\
\indent
11.~A.M.~Kobos, B.A.~Brown, R.~Lindsay and G.R.~Satchler,
"Folding model analysis of elastic and inelastic $\alpha$--particle scattering
using a density dependent force",
{\it Nucl.~Phys.~A\/}, Vol.~425, pp.~205--232, 1984.\\
\indent
12.~H.~Oberhummer and G.~Staudt, "Reaction mechanism for astrophysically
relevant processes", Editor: H.~Oberhummer,
{\it Graduate Texts in Contemporary Physics, Nuclei in the Cosmos\/},
pp.~29--59, Springer Verlag, Heidelberg, 1991.\\
\indent
13.~H.~De Vries, C.W.~De Jager and C.~De Vries, "Nuclear
charge--density--distribution parameters from elastic electron scattering",
At.~Data and Nucl.~Tables, Vol.~36, pp.~495--536, 1987.\\
\indent
14.~Y.K.~Gambhir, P.~Ring and A.~Thimet, "Relativistic mean field
   theory for finite nuclei", {\it Ann.~Phys.~(NY)\/},
   Vol.~198, pp.~132--179, 1990.\\
\indent
15.~P.~Mohr, H.~Abele, R.~Zwiebel, G.~Staudt, H.~Krauss, H.~Oberhummer,
A.~Denker, J.W.~Hammer and G.~Wolf, "Alpha--scattering
and Capture Reactions in the A=7 system at low
energies", {\it Phys.~Rev.~C\/}, Vol.~48, pp.~1420--1427, 1993.\\
\indent
16.~J.~T$\bar{\rm o}$ke and T.~Matulewicz, "An effective
energy--dependent one--body potential for
the low--energy particle--nucleus scattering problem. The one open channel
case"
{\it Act.~Phys.~Polon.~B\/}, Vol.~B14,
pp.609--616, 1983.\\
\indent
17.~T.~Matulwicz, P.~Decowski, M.Kici\'ska-Habior, B.~Sikora,
and J.~T$\bar{\rm o}$ke, " Analysis of the
$^{28}$Si(p,$\gamma$)$^{29}$P reaction data in the
region of the subbarrier single particle resonances",
{\it Act.~Phys.~Polon.~B\/}, Vol.~B14,
pp.617--624, 1983.\\
\indent
18.~M.~Kici\'ska-Habior, ~T.~Matulwicz and P.~Decowski,
"Analysis of the $^{11}$B(p,$\gamma_{0}$)$^{12}$C
reaction in terms of the direct and semi--direct
capture models", {\it Z.~Phys.~B\/}, Vol.~322, pp.611--615, 1983.\\
\indent
19.~T.R.~Donoghue, A.F.~Behof, S.E.~Darden, "Angular distributions
  of neutrons elastically scattered from ${^{18}{\rm O}}$", {\it
Nucl.~Phys.\/},
  Vol.~54, pp.33--48, 1964.\\
\indent
20.~H.~Abele, computer code DFOLD, Universit\"at T\"ubingen, 1991,
unpublished.\\
\indent
21.~B.H.~Wildenthal,  {\it Int.~Symp.~on Nuclear Shell Models\/},
  pp.~360, Singapore, World Scientific, 1985.\\
\indent
22.~D.~Zwarts, "Ritsschil -- a new program for shell--model calculations",
{\it Comp.~Phys.~Com.\/}, Vol.~38, pp.~365--388, 1985.\\
\indent
23.~H.~Herndl, code SPECTROS, TU Wien, 1992, unpublished.\\
\indent
24.~F.~Ajzenberg--Selove, "Energy levels of light nuclei A=18--20",
   {\it Nucl.~Phys.~A\/}, Vol.~475, pp.1--198, 1987.\\
\indent
25.~H.~Krauss, code TEDCA, TU Wien, 1992, unpublished.\\
\indent
26.~H.~Beer, F.~K\"appeler and M.~Wiescher, "Neutron capture on light nuclei
in high entropy scenarios", Editor: J.~Kern,
{\it Proceedings of the 8th International Symposium on Capture Gamma--Ray
Spectroscopy and Related Topics, Fribourg, Switzerland, September 20--24
1993\/},
pp.~756--758, World Scientific, Singapore, 1994.\\
\indent
27.~J.~Meissner, preliminary results, private communication.
\end{document}